\documentstyle[amssymb,12pt]{article} 

\newtheorem{th}{Theorem}[section]
\newtheorem{lm}[th]{Lemma} 
\newtheorem{co}[th]{Corollary}

\newcommand{\ket}[1]{\left | #1 \right \rangle}
\newcommand{\rr}{\mbox{$\Bbb R$}}

\newcommand{\cc}{\mbox{$\Bbb C$}}
\newcommand{\ov}{\overline} 
\newcommand{\vph}{\varphi}
\newcommand{\vep}{\varepsilon} 
\newcommand{\wh}{\widehat}
\newcommand{\sgn}[1]{\mathop{\rm sgn}\left ( #1 \right )}

\begin{document}

\title{\bf On the Structure of Additive Quantum Codes and\\
the Existence of Nonadditive Codes}
\author{Vwani P. Roychowdhury\thanks{e--mail: {\tt vwani@ee.ucla.edu}} 
\hspace{2cm} Farrokh Vatan\thanks{e--mail: {\tt vatan@ee.ucla.edu}} \\
Electrical Engineering Department\\   UCLA\\   Los Angeles, CA 90095}

\date{ }

\maketitle

\begin{abstract}

We first present a useful characterization of additive (stabilizer)
quantum error--correcting codes. Then we present several examples of
nonadditive codes.  We show that there exist infinitely many
non-trivial nonadditive codes with different minimum distances, and
high rates. In fact, we show that nonadditive codes that correct $t$
errors can reach the asymptotic rate $R=1-2H_2(2t/n)$, where $H_2(x)$
is the binary entropy function.  Finally, we introduce the notion of
{\em strongly} nonadditive codes (i.e., quantum codes with the
following property: the trivial code consisting of the entire Hilbert
space is the only additive code that is equivalent to any code
containing the given code), and provide a construction for an ((11,2,
3)) strongly nonadditive code.

\end{abstract}

\section{Introduction}

Almost all quantum error--correcting codes known so far are additive
(or stabilizer) codes.  An additive code can be described as follows.
Consider the group $\cal G$ of unitary operators on the Hilbert space
$\cc^{2^n}$ defined by the tensor products $\pm M_1\otimes M_2\otimes
\cdots \otimes M_n$, where each $M_i$ is either the identity
$I=\pmatrix{1 & 0 \cr 0 & 1 \cr}$ or one the Pauli matrices
$\sigma_x$, $\sigma_z$, or $\sigma_y=\sigma_x\sigma_z$.  Then an
additive code is a subspace $\cal Q$ of $\cc^{2^n}$ for which there is
an Abelian subgroup $H$ of $\cal G$ such that every vector of $\cal Q$ is a
fixed point of every operator in $H$
\cite{orthogeo,gf4,gottesman}. This approach leads to a close connection
between self--orthogonal (under a specific inner product) linear binary codes 
and additive codes, such that the minimum distance of the additive code is
determined from the binary code.

It is natural to ask whether there is any quantum error--correcting
code that can not be constructed in this way, directly or via some
equivalence.  We should make here a comment on the correct formulation
of this question. Since the dimension of every additive quantum code
is a power of 2, any quantum code whose dimension is not a power of 2
is not additive or equivalent to an additive code; specially, any
subspace of an additive code with dimension not a power of 2 is a
nonadditive code.  We call such codes {\em trivial nonadditive codes}.
But we prove a general theorem which shows that
infinite families of non-trivial nonadditive codes with different
values of $d$ exist. The nonadditiveness of these codes does not
follows from their dimensions (the dimensions of these codes are also
powers of two), but from their very special structure. Moreover, we
show that these nonadditive codes asymptotically reach the same rate as
Calderbank--Shor--Steane codes.

We also propose the notion of {\em strongly nonadditive} codes: a
quantum code $\cal Q$ is strongly nonadditive if the trivial code
$\cc^{2^n}$ is the only additive code that contains any code
equivalent to $\cal Q$.  Now the interesting problem is to find
strongly nonadditive quantum codes. Recently in \cite{nonadditive} it
is shown that a $((5,6,2))$ strongly nonadditive code exists, which is
better than any $((5,K,2))$ additive code. Later in \cite{rains_d2},
Rains showed that there exists $((2m,4^{m-1},2))$ nonadditive code,
for all $m\geq 3$.  We present an $((11,2,3))$ strongly nonadditive
code.

In Section \ref{structure} we give a characterization of additive
codes. This characterization is based on the special structure of some
basis of the code, and provides an intuition for constructing the
non-additive codes of Section 4.2.  Finally, in Section
\ref{nonaddsection} first we find a criterion that guarantees
additiveness and strongly nonadditiveness of quantum codes then we
present our example strongly nonadditive code.  Moreover, we give more
examples of nonadditive codes; we conjecture these codes are also
strongly nonadditive.

\section{Preliminaries}
\label{preliminaries}

Consider the Hilbert space $\cc ^{2^n}$ with its standard basis
$\ket{v_1},\ldots , \ket{v_{2^n}}$, where $v_1,\ldots , v_{2^n}$ is a list of
binary vectors of length $n$ in $\{ 0,1\}^n$.  For every binary vector $\alpha$
of length $n$, we define the unitary operators $X_\alpha$ and $Z_\alpha$ by
following equations 
\begin{eqnarray*} 
        X_\alpha \ket{v_i} & = & \ket{v_i+\alpha} ,\\ 
        Z_\alpha \ket{v_i} & = & (-1)^{v_i\cdot \alpha}\ket{v_i} .  
\end{eqnarray*}
Note that $X_\alpha Z_\beta=(-1)^{\alpha \cdot \beta}Z_\beta X_\alpha$.

Let $\cal G$ be the group of all unitary operators of the form
$\pm M_1\otimes\cdots\otimes M_n$, where $M_i\in\{\, I,\sigma_x,\sigma_y,
\sigma_z\,\}$. Then every member of $\cal G$ can be represented uniquely as
$(-1)^\lambda X_\alpha Z_\beta$, where
$\lambda\in \{ 0,1\}$ and $\alpha , \beta \in \{ 0,1 \}^n$.
For every subset $\cal S$ of $\cal G$, let $\ov{\cal S}\subset \{ 0,1
\}^{2n}$ be the set of all vectors $(\alpha | \beta)$ such that either $X_\alpha
Z_\beta \in {\cal S}$ or $-X_\alpha Z_\beta \in {\cal S}$.  We say $\ov{\cal S}$
is {\em totally singular} if for every $(\alpha | \beta) \in \ov{\cal S}$ we
have $\alpha \cdot \beta =0$.  We also define a special inner product on $\{
0,1\}^{2n}$ as \begin{equation} \bigl ( (a|b),(a'|b')\bigr )=a\cdot b'+a'\cdot b
, \label{innerproduct} \end{equation} where the right--hand side is evaluated in
GF(2).  For any quantum code $\cal Q$ in $\cc^{2^n}$, we define the {\em
stabilizer} ${{\cal H}_{\cal Q}}$ of $\cal Q$ as 
\[ {{{\cal H}_{\cal Q}}}=\left \{\, \vph\in {\cal G} :
\vph\ket{x}=\ket{x}\ \mbox{for every $\ket{x}$ in $\cal Q$}\, \right \} .  \]
Then it is easy to check that ${{\cal H}_{\cal Q}}$ is an Abelian group and every element of
${{\cal H}_{\cal Q}}$ squares to the identity operator.  So $\ov{{{\cal H}_{\cal Q}}}$ is totally
singular.  It also follows that ${{\cal H}_{\cal Q}}$ is isomorphic to a vector space
$\mbox{GF}(2)^m$, for some $m$.  This means that ${{\cal H}_{\cal Q}}$ is generated by
operators $\vph_1,\ldots ,\vph_m\in{{{\cal H}_{\cal Q}}}$ and verey $\vph\in {{{\cal H}_{\cal Q}}}$ can be
writen (uniquely, up to the order of the $\vph_i$'s) as
$\vph={\vph_1}^{c_1}\cdots {\vph_m}^{c_m}$, where $c_i\in \{ 0,1\}$.  In this
case the quantum code $\cal Q$ has dimension $2^{n-m}$.  Suppose that
$\vph_i=(-1)^{\lambda _i}X_{\alpha _i}Z_{\beta _i}$.  So $\ov{{{\cal H}_{\cal Q}}}$ can be
determined by its $m\times (2n)$ binary generating matrix 
\begin{equation}
M=\left ( \begin{array}{c|c} \alpha _1 & \beta _1 \\ \vdots & \vdots \\ \alpha_m
& \beta_m \end{array} \right ) .  
\label{Mmatrix} 
\end{equation} 
Note that if
such matrix $M$ obtained from a stabilizer, then $\alpha_i\cdot \beta_i=0$ and
$\alpha_i\cdot \beta_j+\alpha_j\cdot\beta_i=0$, for every $i$ and $j$.  A
quantum code $\cal Q$ is called {\em additive} (or {\em stabilizer}) if it is
defined by its stabilizer ${{\cal H}_{\cal Q}}$, i.e., 
\[ {\cal Q}=\left \{\, \ket{x}\in
\cc^{2^n} :  \vph\ket{x}=\ket{x}\ \mbox{for every $\vph\in{{{\cal H}_{\cal Q}}}$}\, \right \}
.  \]

The quantum codes
${\cal Q}_1$ and ${\cal Q}_2$ in $\cc^{2^n}$ are {\em locally equivalent}
if there is a transversal operator ${\cal U}=u_1\otimes \cdots \otimes u_n$,
with $u_i\in \mbox{SU}(2)$, mapping ${\cal Q}_1$ into ${\cal Q}_2$. We say these
codes are {\em globally equivalent}, or simply equivalent, if ${\cal Q}_1$
is locally equivalent to a code obtained from  ${\cal Q}_2$ by a permutation
on qubits.

A quantum code ${\cal Q}\subseteq \cc^{2^n}$ is called {\bf nonadditive}
if it is not equivalent to any additive code; moreover, $\cal Q$ is
{\bf strongly nonadditive} if the only additive code that contains
any code equivalent to $\cal Q$ is the trivial
code $\cc^{2^n}$; in other words, if $\pm X_\alpha Z_\beta$ is in the stabilizer  
of any code equivalent to a supercode of $\cal Q$ then $\alpha=\beta={\bf 0}$. 

A $K$--dimensional subspace of $\cc^{2^n}$ that as an error--correcting quantum
code can protect against $< d/2$ errors, is called an $((n,K,d))$ code.  If This
code is additive, then $K=2^k$, for some $k$, and it is called an $[[n,k,d]]$
code.  The following theorem gives a sufficient condition that a subspace of
$\cc^{2^n}$ to be an $((n,K,d))$ code.  Here $\mbox{wt}(c)$ denotes the Hamming
weight of the binary vector $c$, i.e.  the number of 1--components of $c$, and
$\alpha \cup \beta$ is the binary vector result of componentwise OR operation of
$\alpha$ and $\beta$; for example $(10110)\cup (00101)=(10111)$.

\begin{th} 
{\em (\cite{bdsw}, \cite{knilllaf})} Let $\cal Q$ be a
$K$--dimensional subspace of $\cc^{2^n}$.  Consider an orthonormal basis for
$\cal Q$ of the form $\left \{\, \ket{c_i} :  i=1,\ldots ,K\, \right \}$.  Then
$\cal Q$ is an $((n,K,d))$ code if $\langle c_i\, | \, X_\alpha Z_\beta\, | \,
c_j\rangle =0$ for every $1\leq i,j\leq K$ and for every $\alpha ,\beta\in \{
0,1\}^n$ with $1\leq \mbox{\em wt}(\alpha \cup \beta )\leq d-1$.  In general, a
necessary and sufficient condition for $\cal Q$ to be an $((n,K,d))$ code is
that for all $1\leq i,j \leq K$ and $\mbox{\em wt}(\alpha \cup \beta )\leq d-1$
we have $\langle c_i\, | \, X_\alpha Z_\beta\, | \, c_i\rangle = \langle c_j\, |
\, X_\alpha Z_\beta\, | \, c_j\rangle $ and if $i\neq j$ then $\langle c_i\, |
\, X_\alpha Z_\beta\, | \, c_j\rangle =0$.  
\label{condition} 
\end{th}

For an additive code $\cal Q$ with stabilizer ${{\cal H}_{\cal Q}}$ there is a sufficient
condition in term of the dual of ${{\cal H}_{\cal Q}}$ with respect to the inner product
defined by equation (\ref{innerproduct}) for $\cal Q$ to be a
$t$--error--correcting code.

\begin{th} 
{\em (\cite{orthogeo}, \cite{gottesman})} Let $\cal Q$ be an additive
code with stabilizer ${{\cal H}_{\cal Q}}$.  Let $\ov{{{\cal H}_{\cal Q}}}^\perp$ be the space orthogonal
to $\ov{{{\cal H}_{\cal Q}}}$ with respect to the inner product {\em (\ref{innerproduct})}.
If for every binary vectors $\alpha , \beta\in\{ 0,1\}^n$ with $\mbox{\em
wt}(\alpha \cup \beta )\leq d-1$ we have $(\alpha | \beta )\not \in 
\ov{{{\cal H}_{\cal Q}}}^\perp \setminus \ov{{{\cal H}_{\cal Q}}}$ then 
$\cal Q$ is an $[[n,k,d]]$.
\label{stabilizertheo} 
\end{th}

\section{The structure of additive codes}
\label{structure}

We give a characterization of additive quantum error--correcting codes.  Suppose 
that the matrix $M$ in (\ref{Mmatrix}) specifies the stabilizer of an additive code
$\cal Q$.  If we add one row of $M$ to another row of it, the resulting matrix
also generates $\ov{{{\cal H}_{\cal Q}}}$; i.e., the new matrix can be obtained from some
other basis of ${{\cal H}_{\cal Q}}$.  So we can assume, without loss of generality, that $M$
has the following structure:  
\begin{equation} M = \left ( \begin{array}{c|c} 
          A & B \\ \hline 0 & P \end{array} \right ) 
          =\left ( \begin{array}{c|c} 
          a_1 & b_1 \\ \vdots & \vdots \\ a_r & b_r \\ \hline 0 & P 
                              \end{array} \right ) ,
\label{mabp} 
\end{equation} 
where $A$ and $P$ are full-rank matrices, and $A$ is
a generator matrix for the binary code $\cal C$.

The Calderbank--Shor--Steane (CSS) codes are special class of additive codes with a 
simple structure. In this section we show that the structure of any additive code is 
similar to the structure of CSS codes with some differences. Let us first explain the
construction of theses codes.

Suppose that $\cal C$ is a weakly self--dual
$[n,k,d_0]$ binary code (i.e., ${\cal C}\subseteq {\cal C}^\perp$). Suppose that
$\mbox{dist}({\cal C}^\perp)\geq d$. The vectors
$\displaystyle \ket{x_a}=\sum_{c\in{\cal C}}\ket{c+a}$, where $a\in{\cal C}^\perp$, 
form the CSS code $\cal Q$. (To simplify the notation, throughout this paper we
delete the normalization factors.)
Then $\cal Q$ is an $[[n,n-2k,d]]$ additive code.
For $a,a'\in{\cal C}^\perp$, we have $\ket{x_a}=\ket{x_{a'}}$ if and only if $a$ and $a'$
belong to the same coset of $\cal C$ in ${\cal C}^\perp$; so the dimension of $\cal Q$
is equal to the number of cosets of $\cal C$ in ${\cal C}^\perp$, which is $2^{n-2k}$.

We show that for any additive code we have a similar basis, but here we have to add
some ``signs'' to the states; i.e., the basis consists of vectors of the form
$\displaystyle \ket{x_a}=\sum_{c\in{\cal C}}\sgn{c+a}\ket{c+a}$, where 
$\cal C$ is some binary linear code, $a$'s belong to some other linear code
(not necessarily ${\cal C}^\perp$) and 
$\sgn{c+a}$'s are chosen in a very special way from $\{ -1,+1\}$ (see equations
(\ref{sgn4}) and (\ref{sgn5})). Moreover, we show that these bases characterize
additive codes, in the sense that any quantum code that has such a basis (with signs
$\sgn{c+a}$'s satisfying the equations detemined in the following theorems) is additive.

\begin{th} 
Suppose that the $2^{n-m}$--dimensional space ${\cal Q}\subseteq
\cc^{2^n}$ is an additive quantum error--correcting code with stabilizer 
${{\cal H}_{\cal Q}}$.  Suppose that the full-rank matrix $M$ in {\em (\ref{mabp})} 
generates $\ov{{{\cal H}_{\cal Q}}}$; i.e., 
$a_i\cdot b_i=0$ and $a_i\cdot b_j+a_j\cdot b_i=0$, for all $1\leq
i,j \leq r$, and $a_i$'s belong to the dual space of $P$.  More specifically,
let ${{\cal H}_{\cal Q}}$ be generated by $\bigl \{ \vph_1,\ldots , \vph _m\bigr \}$, where
$\vph_i=\vep_iX_{a_i}Z_{b_i}$, for some $\vep_i\in\{ -1,+1\}$ and $a_i=\mbox{\bf
0}$ for $r < i\leq m$.  Let $\cal C$ be the the binary linear code generated by
$\bigl \{ a_1,\ldots ,a_r \bigr \}$.  Then there are independent binary vectors
$\gamma_1,\ldots , \gamma_{n-m}$ in $\{ 0,1\} ^{n}\setminus {\cal C}$ generating
the linear space $\Gamma$ such that the followings hold.

{\em (i)} $\cal Q$ has a basis consists of the vectors of the form
\begin{equation} 
\ket{x_\gamma} =\sum_{c\in{\cal C}}
\sgn{c+\gamma}\ket{c+\gamma} , \qquad \gamma \in \Gamma, 
\label{qbasis}
\end{equation} 
for some $\sgn{c+\gamma}\in \{ -1 , +1\}$.

{\em (ii)} $\sgn{c+\gamma}$'s satisfy the following identities:
\begin{equation} 
\sgn{\gamma}=1\qquad \mbox{for}\ \gamma\in\Gamma ,
\label{sgn1a} 
\end{equation} 
\begin{equation} 
\sgn{a_i}=\vep_i\qquad \mbox{for}\ 1\leq i\leq r, 
\label{sgn1b} 
\end{equation} 
\begin{equation}
\sgn{\sum_{j=1}^\ell a_{i_j}} = (-1)^{b_{i_1}\cdot\sum_{j=2}^\ell a_{i_j}}
(-1)^{b_{i_2}\cdot\sum_{j=3}^\ell a_{i_j}} \cdots (-1)^{b_{i_{\ell-1}}\cdot
a_{i_\ell}}\vep_{i_1}\cdots \vep_{i_\ell} , 
\label{sgn2} 
\end{equation}
\begin{equation} 
\sgn{\sum_{j=1}^\ell a_{i_j}+\gamma} = (-1)^{\gamma \cdot
\sum_{j=1}^\ell b_{i_j}} \sgn{\sum_{j=1}^\ell a_{i_j}} , \qquad\mbox{for every}\
\ell\geq 1\ \mbox{and}\ \gamma\in\Gamma .  
\label{sgn3} 
\end{equation}
\label{sgntheorem1} 
\end{th}

{\bf Proof.}  (i) Let $\cal D$ be the space of vectors in $\{0,1\}^n$ orthogonal to 
the rows of
$P$.  Then the dimension of $\cal D$ is $n-m+r$ and ${\cal C}\subseteq {\cal
D}$.  Choose vectors $\gamma_1,\ldots ,\gamma_{n-m}$ such that $\left \{
a_1,\ldots ,a_r,\gamma_1,\ldots ,\gamma_{n-m}\right \}$ be a basis for $\cal D$.
Let $\Gamma$ be the space generated by $\{\, \gamma_1,\ldots,\gamma_{n-m}\,\}$.
There are $2^{n-m+r}/ 2^{r}=2^{n-m}$ cosets of $\cal C$ in $\cal D$; each coset
can be represented as $\gamma +{\cal C}$ where $\gamma\in\Gamma$ is a linear
combination of $\gamma_j$'s.  It is easy to check that in fact $\displaystyle
\ket{x_\gamma}=\sum_{\vph\in{{{\cal H}_{\cal Q}}}}\vph \ket{\gamma}$, because each operator in
${{\cal H}_{\cal Q}}$ can be written as $\pm X_\alpha Z_\beta$, where $\alpha\in{\cal C}$
and $\beta$ is in the group generated by $b_1,\ldots,b_r$ plus the rows of $P$. 
So, for every
$\psi\in{{{\cal H}_{\cal Q}}}$,
\[ \psi\ket{x_\gamma}=\sum_{\vph\in{{{\cal H}_{\cal Q}}}}\psi\vph\ket{\gamma}
=\sum_{\vph\in{{{\cal H}_{\cal Q}}}}\vph\ket{\gamma} =\ket{x_\gamma} .\]
Therefore, $\ket{x_\gamma}\in{\cal Q}$.  On the other hand, $\ket{x_\gamma}$ and
$\ket{x_{\gamma '}}$ are orthogonal for $\gamma \neq \gamma '$.  So the
$2^{n-m}$ vectors $\ket{x_\gamma}$ form a basis for $\cal Q$.

(ii) Condition (\ref{sgn1a}) follows form the fact that $I\ket{\gamma}=
\ket{\gamma}=\sgn{\gamma}\ket{\gamma}$, and (\ref{sgn1b}) follows from the fact that
$\vep_iX_{a_i}Z_{b_i}\ket{\mbox{\bf 0}}=\vep_i\ket{a_i}$ should be equal to
$\sgn{a_i}\ket{a_i}$.

We can prove (\ref{sgn2}) by an induction on $\ell$.  For $\ell =1$, it reduces
to (\ref{sgn1b}).  Suppose that (\ref{sgn2}) is true for $\ell$.  Then 
(here we are using the fact that $a_i\cdot b_i=0$)
\[    \vep_{i_1}X_{a_{i_1}}Z_{b_{i_1}}\sgn{\sum_{j=1}^{\ell+1}a_{i_j}}
\ket{\sum_{j=1}^{\ell+1}a_{i_j}}  =  \vep_{i_1}(-1)^{b_{i_1}\cdot\left ( \sum_
{j=2}^ {\ell+1}a_{i_j}\right )}\sgn{\sum_{j=1}^{\ell+1}a_{i_j}} \ket{\sum_{j=2}^
{\ell+1}a_{i_j}}         \] 
should be equal to 
\[   \sgn{\sum_{j=2}^{\ell+1}a_{i_j}} \ket{\sum_{j=2}^{\ell+1}a_{i_j}} , \]
so it follows 
\[ \sgn{\sum_{j=1}^{\ell+1}a_{i_j}}=\vep_{i_1}(-1)^{b_{i_1}\cdot \left ( \sum
_{j=2}^{\ell+1}a_{i_j}\right )}\sgn{\sum_{j=2}^{\ell+1}a_{i_j}} .\] 
Then the induction hypothesis implies (\ref{sgn2}).

By a similar inductive argument (\ref{sgn3}) can be proved.  $\blacksquare$

\vspace{8mm} 
In the next theorem we present relations among $\sgn{c+\gamma}$'s
which characterize the additive codes.

\begin{th} 
Every sign $\sgn{c+\gamma}$ in Theorem {\em \ref{sgntheorem1}} is a
function of the following signs 
\[ \sgn{a_i},\ \sgn{a_i+a_j}\ \mbox{and}\
\sgn{a_i+\gamma_k}\qquad \mbox{for}\ 1\leq i,j\leq r\ \mbox{and}\ 
1\leq k\leq n-m .\] 
More specifically, the following relations hold.  For every nonempty
subsets $S\subseteq \{ 1,2,\ldots , r\}$ and $T\subseteq\{ 1,2,\ldots ,n-m\}$ we
have 
\begin{equation} 
\sgn{\sum_{i\in S}a_i}=\prod_{i\in S}\left [ \mbox{\em
sgn}(a_i)\right ]^{|S|} \prod_{ i<j\atop i,j\in S} \sgn{a_i+a_j} , 
\label{sgn4} 
\end{equation} 
\begin{equation}
\sgn{\sum_{i\in S}a_i+\sum_{j\in T}\gamma_j} = \sgn{\sum_{i\in S}a_i}
\left [ \prod_{i\in S}\sgn{a_i}\right ] ^{|T|}
\prod_{i\in S \atop j\in T} \sgn{a_i+\gamma_j}.  
\label{sgn5} 
\end{equation} 
\label{sgntheorem2} 
\end{th}

{\bf Proof.}  From (\ref{sgn1b}) and (\ref{sgn2}) it follows 
\begin{equation} 
             (-1)^{b_i\cdot a_j}=\sgn{a_i}\sgn{a_j}\sgn{a_i+a_j} .  
\label{sgn6} 
\end{equation} 
Now (\ref{sgn4}) follows from (\ref{sgn1b}) and (\ref{sgn2}) by expanding the 
inner products and substituting $(-1)^{b_i\cdot a_j}$ from (\ref{sgn6}).

Similarly, from (\ref{sgn3}) it follows 
\begin{equation} 
              (-1)^{b_i\cdot \gamma_j}=\sgn{a_i}\sgn{a_i+\gamma_j} .  
\label{sgn7} 
\end{equation}
Then (\ref{sgn3}) implies (\ref{sgn5}).  $\blacksquare$

\vspace{8mm} 
Now we give a characterization of additive codes.

\begin{th} 
Let ${\cal Q}$, a $2^{n-m}$--dimensional subspace of $\cc^{2^n}$, be
a quantum error--correcting code.  Suppose that there is
a linear binary code ${\cal C}\subseteq \{ 0,1\}^n$ with basis
$\{\, a_1,\ldots,a_r\,\}$, $r\leq m$, and vectors 
$\gamma_1\ldots ,\gamma_{n-m}$ with the property that $\{\, a_1,\ldots,a_r,\gamma_1
\ldots ,\gamma_{n-m}\,\}$ is an indepentdent set ($\gamma_i$'s are basis for 
some binary code $\Gamma$). 
Then $\cal Q$ is an additive code if $\cal Q$ has a basis $\cal B$ of
the form {\em (\ref{qbasis})} where the signs $\sgn{c+\gamma}$ satisfy equations
{\em (\ref{sgn1a})}, {\em (\ref{sgn4})} and {\em (\ref{sgn5})}.
\label{charactheorem} 
\end{th}

{\bf Proof.}  Suppose that $a_1,\ldots ,a_r$ is a basis for the binary code $\cal C$.  
If $r<m$ then
let $P$ be a generator matrix for the linear code that is orthogonal to both
$\cal C$ and $\Gamma$.  Let $p_1,\ldots ,p_{m-r}$ be the rows of $P$.

For $1\leq i\leq r$, let $b_i\in \{ 0,1\}^n$ be any vector that satisfies the
equations 
\begin{eqnarray*}
       (-1)^{b_i\cdot a_j} & = & \sgn{a_i}\sgn{a_j}\sgn{a_i+a_j} , \quad 
                                   1\leq j \leq r , \\ 
  (-1)^{b_i\cdot \gamma_j} & = & \sgn{a_i}\sgn{a_i+\gamma_j} ,\quad 
                                   1\leq j\leq n-m .
\end{eqnarray*} 
Such $b_i$ exists, because the above eqautions can be written as
a system of $n-m+r$ linear equations with independent vectors $a_j$'s and
$\gamma_j$'s as its coefficient vectors.  Consider the group ${{\cal H}_{\cal Q}}$ of unitary
operators generated by 
\[ e_i=\sgn{a_i}X_{a_i}Z_{b_i},\quad 1\leq i\leq r,\quad
\mbox{and}\quad f_i=Z_{p_i},\quad 1\leq i\leq m-r .  \] 
(Of course we consider $f_i$'s only if $r<m$.)  Then ${{\cal H}_{\cal Q}}$ is
Abelian:  $e_ie_j=e_je_i$ (for $i\neq j$) follows from the fact that
$(-1)^{b_i\cdot a_j}=(-1)^{b_j\cdot a_i}=\sgn{a_i}\sgn{a_j} \sgn{a_i+a_j}$;
$e_if_j=f_je_i$ and $f_if_j=f_jf_i$ are obvious.  Also every element of ${{\cal H}_{\cal Q}}$
sqaures to identity:  ${e_i}^2=I$ follows from the fact that $(-1)^{a_i\cdot
b_i}=\sgn{a_i}\sgn{a_i}\sgn{a_i+a_i}=1$ so $a_i\cdot b_i=0$; ${f_i}^2=I$ is
obvious.  Thus ${{\cal H}_{\cal Q}}$ is the stabilizer of an additive quantum code $\cal Q'$
of dimension $2^{n-m}$.  Consider the basis $\cal B'$ for $\cal Q'$ provided by
Theorem \ref{sgntheorem1}.  Then, by Theorem \ref{sgntheorem2}, ${\cal B}={\cal
B}'$.  So ${\cal Q}={\cal Q}'$, and $\cal Q$ is an additive code.
$\blacksquare$

\section{Existence of nonadditive codes}
\label{nonaddsection}

\subsection{Quantum codes equivalent to additive codes}
\label{equisubsection}

We study the quantum codes equivalent to additive codes. For such code
$\cal Q$, we find a sufficient condition that guarantees that
the stabilizer of $\cal Q$ contains a nontrivial operator.

We begin with some useful notions and notations.
Let $\ket{c_1},\ldots ,\ket{c_{2^n}}$ be the standard orthonormal basis of
$\cc^{2^n}$, where each $c_i$ is a binary vector of length $n$. For the
vector $\displaystyle \ket{x}=\sum_{i=1}^{2^n} \lambda _i \ket{c_i}$, we
define the {\em support} of $\ket{x}$ as
\[\mbox{supp}(\ket{x})=\left\{\, c_i\in\{0,1\}^n:\lambda_i\neq 0\,\right\} .\]

Let ${\cal C}\subseteq \{ 0,1\}^n$ be a set of binary vectors. Define the
vector $\ket{\cal C}$ in $\cc^{2^n}$ as
\[ \ket{\cal C}={1\over |{\cal C}|^{1/2}}\sum_{c\in {\cal C}}\ket{c} . \]
(If $\cal C$ is empty then $\ket{\cal C}$ is the zero vector.) For any
binary vector $\alpha$ of length $m<n$, define
\begin{equation}
   {\cal C}_{\alpha}=\left \{\, x\in\{ 0,1\}^{n-m} : (\alpha,x)\in{\cal C}
            \, \right \} .
\label{calpha}
\end{equation}
So to construct ${\cal C}_{\alpha}$, consider all vectors in $\cal C$
starting with $\alpha$ (if there is any), then delete $\alpha$ from these
vectors. Note that ${\cal C}_{\alpha}$ may be empty.

For a quantum code $\cal Q$, let us define {\em the generalized stabilizer}
of $\cal Q$ as the set $GS({\cal Q})$ of all unitary operators $\cal V$ on
$\cc^{2^n}$ such that ${\cal V}\ket{x}=\ket{x}$ for every $\ket{x}\in{\cal Q}$.
Then the {\em stabilizer} of $\cal Q$ is $\mbox{St}({\cal Q})={\cal G}\cap
GS({\cal Q})$.

\begin{lm}
Suppose that the quantum codes ${\cal Q}_1$ and ${\cal Q}_2$ are locally equivalent
via the transversal unitary operator $\cal U$. Then for every $M\in
GS({\cal Q}_1)$ the operator ${\cal U}M{\cal U}^\dagger$ is in $GS({\cal Q}_2)$.
\label{umudagger}
\end{lm}

{\bf Proof.} Let $\ket{x}\in{\cal Q}_2$. There is $\ket{y}\in{\cal Q}_1$ such
that $\ket{x}={\cal U}\ket{y}$.  Since $M\ket{y}=\ket{y}$, so
$(M{\cal U}^\dagger){\cal U}\ket{y}=\ket{y}$, and therefore
$({\cal U}M{\cal U}^\dagger){\cal U}\ket{y}={\cal U}\ket{y}$. This implies
$({\cal U}M{\cal U}^\dagger)\ket{x}=\ket{x}$. $\blacksquare$

\vspace{8mm}
We are interested in the case of $M\in{\cal G}$, i.e., $M=M_1\otimes\cdots
\otimes M_n$, where $M_j\in\{ I,\sigma_x,\sigma_y,\sigma_z\}$. We define
$\mbox{wt}(M)$ the {\em weight} of any $M\in{\cal G}$ as the number of $j$'s
such that $M_j\neq I$.
In this case ${\cal U}M{\cal U}^\dagger =v_1\otimes \cdots \otimes v_n$ such
that $\det (v_j)=\pm 1$ and if $M_j=I$ then $v_j=I$, otherwise
\begin{equation}
    v_j=\eta_j\pmatrix{a_j & b_j \cr {\pm b_j}^* & -a_j \cr},
    \qquad \eta_j\in \{ 1,i\},\ a_j\in\rr\ \mbox{and}\ b_j\in\cc .
\label{vmatrix}
\end{equation}
If ${\cal U}\in\mbox{SU}(2)^{\otimes n}$ then $\cal U$ is of the form $u_1\otimes
\cdots\otimes u_n$, where each $u_j$ is defined by a matrix of the form
\begin{equation}
 \pmatrix{e^{i\alpha}\cos\theta   &   e^{i\beta}\sin\theta \cr
          -e^{-i\beta}\sin\theta &   e^{-i\alpha}\cos\theta \cr} .
\label{umatrix}
\end{equation}
If $M_j=\sigma_x$, $\sigma_z$ or $\sigma_y$, then the corresponding $v_j$,
repectively, is
\begin{equation}
\left .
\begin{array}{c}
   \pmatrix{\sin 2\theta\cos (\alpha -\beta) & \cos^2\theta e^{i2\alpha}-
             \sin^2\theta e^{i2\beta} \cr
             \cos^2\theta e^{-i2\alpha}-\sin^2\theta e^{-i2\beta} &
             -\sin 2\theta\cos (\alpha -\beta) \cr}, \\  \\
   \pmatrix{\cos 2\theta & -\sin 2\theta e^{i(\alpha +\beta)} \cr
             -\sin 2\theta e^{-i(\alpha +\beta)} & -\cos 2\theta \cr}, \\  \\
             \mbox{or}\
   \pmatrix{-i\sin 2\theta\sin (\alpha -\beta) & -\cos^2\theta e^{i2\alpha}-
             \sin^2\theta e^{i2\beta} \cr
             \cos^2\theta e^{-i2\alpha}+\sin^2\theta e^{-i2\beta} &
             i\sin 2\theta\sin (\alpha -\beta) \cr} .
\end{array}
\right \}
\label{umumatrix}
\end{equation}

We call a matrix $v_i$ as (\ref{vmatrix}) {\em full} if $a_i\cdot b_i\neq 0$;
and we say the unitary operator ${\cal V}=v_1\otimes\cdots\otimes v_n$ 
is {\em thin} if none of $v_i$'s is full. In the next proof we will use this
property that if $\cal V$ is thin then $|\mbox{supp}({\cal V}\ket{x})|=
|\mbox{supp}(\ket{x})|$, for every $\ket{x}$.

A quantum code $\cal Q$ is called {\em real} if $\cal Q$ has a basis consisting
of real vectors; i.e., if $\displaystyle\ket{x}=\sum_{i=1}^{2^n}\lambda_i
\ket{c_i}$ is any vector in the basis, then $\lambda_i\in\rr$, for
every $i$.

An $(n,K,d)$ binary code is a set ${\cal C}\subseteq \{ 0,1\}^n$ of size
$K$ such that any two vectors in $\cal C$ differ in at least $d$ places,
and $d$ is the largest number with this property. Note that an $[n,k,d]$
binary linear code is an $(n,2^k,d)$ binary code.

\begin{th}
Suppose that the quantum codes ${\cal Q}_1$ and ${\cal Q}_2$ are locally
equivalent via the transversal operator $\cal U$, ${\cal Q}_2$ is real 
and ${\cal Q}_2$ contains $\ket{\cal C}$, where $\cal C$ is an
$(n,K,d)$ binary code with $d>k=\lceil\log_2 K\rceil$. Then the 
following claims hold.

(i) The image of $\mbox{\em St}({\cal Q}_1)$ under the mapping
$M\mapsto {\cal U}M{\cal U}^\dagger$, which we call $\Gamma$, 
consists only of unitary operators 
$\pm X_{\alpha}T$, where $T$ is a $Z$--type unitary operator of the form
\begin{equation}
 T=\bigotimes _{j=1}^n \pmatrix{ e^{i\theta_j} & 0 \cr 0 & \pm 
    e^{-i\theta_j}\cr} .
\label{tequ}
\end{equation}

(ii) Let $\Delta =\{\, \alpha\in\{0,1\}^n : \pm X_\alpha T\in\Gamma
\ \mbox{\em for some $T$ of the form (\ref{tequ})} \,\}$.
Suppose that $\mbox{\em St}({\cal Q}_2)$ does not contain any operator of 
the form $\pm X_{\bf 0}Z_\beta$, with $\beta\neq {\bf 0}$.
Then $|\mbox{\em St}({\cal Q}_1)|\leq |\Delta|$.
\label{equiadditheo}
\end{th}

{\bf Proof.} By Lemma \ref{umudagger}, there are $v_i\in \mbox{SU}(2)$,
$1\leq i\leq n$, such that $v_i=I$ or $v_i$ satisfies (\ref{vmatrix}) 
(or, equivalently (\ref{umumatrix})) and for ${\cal V}=v_1\otimes\cdots\otimes v_n$
we have
\begin{equation}
{\cal V}\ket{\cal C}=\ket{\cal C} ,
\label{v1}
\end{equation}
We claim $\cal V$ is a thin
operator. By contradiction, assume $\cal V$ is not thin; and, w.l.o.g.,
$v_1$ is full. Let ${\cal V}_1=v_2\otimes\cdots\otimes v_n$. Define
${\cal C}_0$ and ${\cal C}_1$ as (\ref{calpha}), i.e.,
\[ {\cal C}_0=\left\{\, x\in\{0,1\}^{n-1} : (0,x)\in{\cal C}\,\right\} ,\]
and a similar eqaution for ${\cal C}_1$.  Thus, $\ket{\cal C}=\ket{0}\otimes\ket{{\cal C}_0}+
\ket{1}\otimes\ket{{\cal C}_1}$. Then (\ref{v1})
implies
\begin{eqnarray*}
 a_1 {\cal V}_1\ket{{\cal C}_0}\pm {b_1}^*{\cal V}_1\ket{{\cal C}_1}
         & = & \ket{{\cal C}_0}, \\
 b_1 {\cal V}_1\ket{{\cal C}_0}-a_1{\cal V}_1\ket{{\cal C}_1}
         & = & \ket{{\cal C}_1}.
\end{eqnarray*}
This shows that ${\cal C}_0$ and ${\cal C}_1$
both should be non--empty. By solving this system, we get
\begin{eqnarray*}
       {\cal V}_1\ket{{\cal C}_0} & = & -a_1\ket{{\cal C}_0}\mp
                                        {b_1}^*\ket{{\cal C}_1} , \\
       {\cal V}_1\ket{{\cal C}_1} & = & -b_1\ket{{\cal C}_0}+
                                        a_1\ket{{\cal C}_1} .
\end{eqnarray*}
If ${\cal V}_1$ is thin then 
$|\mbox{supp}(\ket{{\cal C}_0})|=|\mbox{supp}(\ket{{\cal C}_0})|$, but since
$\mbox{supp}(\ket{{\cal C}_0})\cap\mbox{supp}(\ket{{\cal C}_1})=
\emptyset$,  it follows that ${\cal V}_1$ is not thin and
for some $i$, $2\leq i\leq n$, $v_i$ should be a full matrix.
Assume, w.l.o.g., $v_2$ is full. Then, with a similar calculation for
${\cal V}_2=v_3\otimes\cdots\otimes v_n$,
\[ {\cal V}_2\ket{{\cal C}_{\alpha_i}}=\sum_{j=1}^4 \lambda_j
              \ket{{\cal C}_{\alpha_j}}, \qquad 1\leq i\leq 4, \]
where $\alpha_i$ is a binary vector of length 2 and each $\lambda _j$ is
a product of entries of $v_1$ and $v_2$ (so each $\lambda _j$ is nonzero).
If $K\geq 4$, then $\mbox{supp}(\ket{{\cal C}_{\alpha_i}})$ are disjoint
(because $d>k$) and they should be non--empty. Therefore, at least one
of $v_3,\ldots ,v_n$ should be full. Again, w.l.o.g., we acn assume $v_3$ is
full. By continuing this argument, we find out that $k$
of $v_i$'s, say
$v_1,\ldots ,v_k$, are full and for ${\cal V}_k=v_{k+1}\otimes\cdots\otimes
v_n$ and any $\beta\in\{ 0,1\}^k$ we have
\begin{equation}
     {\cal V}_k \ket{{\cal C}_{\beta}}
   =\sum_{\alpha\in\{ 0,1\}^k} \lambda_\alpha \ket{{\cal C}_\alpha},
\label{alpha0}
\end{equation}
where each $\lambda_\alpha$ is a product of the entries of $v_1,\ldots ,v_k$,
so all $\lambda_\alpha$ are nonzero. Since $d>k$, all ${\cal C}_\alpha$,
$\alpha\in \{ 0,1\}^k$, have disjoint support. Therefore, for every $\alpha\in
\{ 0,1\}^k$, ${\cal V}_k \ket{{\cal C}_\alpha}\neq 0$. This implies that
for every $\alpha$, the size of $\mbox{supp}(\ket{{\cal C}_\alpha})$ is one.
Therefore, for every $\alpha\in\{0,1\}^k$, either $\ket{{\cal C}_{\alpha 0}}=0$
or $\ket{{\cal C}_{\alpha 1}}=0$
We conclude that ${\cal V}_k$ can not be thin, so at least one of $v_{k+1},
\ldots, v_n$ is full. Suppose that $v_{k+1}$ is full and let ${\cal V}_{k+1}=
v_{k+2}\otimes\cdots\otimes v_n$. Consider any $\beta\in\{ 0,1\}^k$. Then
either $\mbox{supp}(\ket{{\cal C}_{\beta 0}})=\emptyset$ or
$\mbox{supp}(\ket{{\cal C}_{\beta 1}})=\emptyset$. Assume, w.l.o.g.,
that $\mbox{supp}(\ket{{\cal C}_{\beta 1}})=\emptyset$. Therefore, 
$\ket{{\cal C}_{\beta}}=\ket{0}\otimes\ket{{\cal C}_{\beta 0}}$.
Then (\ref{alpha0}) implies
\[ a_{k+1}\ket{0}\otimes {\cal V}_{k+1} \ket{{\cal C}_{\beta 0}}+
   b_{k+1}\ket{1}\otimes {\cal V}_{k+1} \ket{{\cal C}_{\beta 0}}=
   \ket{0}\otimes \sum_{\alpha\in\{ 0,1\}^k} \lambda_\alpha
             \ket{{\cal C}_{\alpha 0}} +
   \ket{1}\otimes \sum_{\alpha\in\{ 0,1\}^k} \lambda_\alpha
             \ket{{\cal C}_{\alpha 1}} . \]
Thus
\[ {\cal V}_{k+1} \ket{{\cal C}_{\beta 0}}={1\over a_{k+1}}
      \sum_{\alpha\in\{ 0,1\}^k} \lambda_\alpha \ket{{\cal C}_{\alpha 0}} =
   {1\over b_{k+1}}\sum_{\alpha\in\{ 0,1\}^k} \lambda_\alpha \ket{{\cal C}
          _{\alpha 1}}   .\]
Therefre
\[ {1\over a_{k+1}}
    \sum_{\alpha\in\{ 0,1\}^k} \lambda_\alpha \ket{{\cal C}_{\alpha 0}}
   -{1\over b_{k+1}}
    \sum_{\alpha\in\{ 0,1\}^k} \lambda_\alpha \ket{{\cal C}_{\alpha 1}} =0. \]
Which is not possible, because in this equation $2^k$ vectors are zero and
the other $2^k$ vectors are linearly independent and all coefficients are
nonzero.

Now to see that the statement (i) of the theorem holds, it is enough to note that
\[ \pmatrix{0 & e^{i\theta}\cr \pm e^{-i\theta} & 0\cr} = \pmatrix{0 & 1\cr 1 & 0\cr }
                          \pmatrix{\pm e^{i\theta} & 0\cr 0 & e^{-i\theta} \cr} .\]

Now we are ready to prove (ii). Suppose that $X_{\alpha_1}Z_{\beta_1}$ and 
$X_{\alpha_2}Z_{\beta_2}$ are in $\mbox{St}({\cal Q}_1)$ and 
$(\alpha_1 ,\beta_1)\neq (\alpha_2,\beta_2)$. Suppose that $X_{\alpha_j}Z_{\beta_j}$
is mapped to ${\cal V}_j=\pm v^j_1\otimes\cdots\otimes v^j_n$, $j=1,2$, 
where each $v_l^j$ is of
the form (\ref{vmatrix}), or more explicitly of the form (\ref{umumatrix}).
Let ${\cal V}_j=X_{a_j}T_j$, $j=1,2$. We assume $a_1=a_2=a$ and derive a contradiction.
Without loss of generality, we can assume $a=(\overbrace{\mathstrut 1,\ldots,1}
^{m\ {\rm times}},0,\ldots ,0)$. Therefore, $v^1_\ell=v^2_\ell=\{ I,\sigma_z\}$, for 
$\ell=m+1,\ldots ,n$; and the matrix of $v^j_\ell$, $j=1,2$ and $\ell=1,\ldots ,m$, is 
anti--diagonal, i.e., it is of the form $\pmatrix{0 & x\cr y & 0\cr}$.

Before we continue
note that the matrices of $v_x=u\sigma_x u^\dagger$, $v_y=u\sigma_y u^\dagger$,
$v_z=u\sigma_z u^\dagger$,
for a fixed $u\in \mbox{SU}(2)$, 
are of the form (\ref{umumatrix}), and if two of $\{\, v_x,v_y,v_z\,\}$ are
anti--diagonal then the third is diagonal, and if one of them is diagonal then the other
two are anti--diagonal. 

Now we show that the operator $X_{\alpha_1+\alpha_2}Z_{\beta_1+\beta_2}$
in $\mbox{St}({\cal Q}_1)$
is mapped to an operator $v_1\otimes\cdots\otimes v_n$ of the form $X_{\bf 0}Z_\beta$ with
$\beta\neq {\bf 0}$, which is the desired contradiction. 
Note that if $v^j_\ell=u_\ell\sigma^ju^\dagger_\ell$, for $j=1,2$ and $\sigma^j\in
\{I,\sigma_x,
\sigma_y,\sigma_z\}$, then $v_\ell=u_\ell\sigma^1\sigma^2u^\dagger_\ell$. For $\ell=m+1,
\ldots ,n$,
since $v^1_\ell$ and $v^2_\ell$ both have diagonal matrices, then either $\sigma^1$ and 
$\sigma^2$ are identical or one of them is the identity operator. In either case 
$v_\ell=I$ or $\sigma_z$.
Similarly, for $i=1,\ldots,m$, $v^1_\ell$ and $v^2_\ell$ both have anti--diagonal matrices
and $v_\ell$ should be either identity or $\sigma_z$. This shows that $v_1\otimes\cdots
\otimes v_n=X_{\bf 0}Z_\beta$. It remains to show that at least 
one of $v_\ell$ is not identity. Since $(\alpha_1+\alpha_2\mid \beta_1+\beta_2)\neq {\bf 0}$,
at least one of $v_\ell$ is of the form $u_l\sigma u_i^\dagger$, where $\sigma\in\{\sigma_x,
\sigma_y,\sigma_z\}$. So the matrix of $v_\ell$ is of the form (\ref{umumatrix}) which
is never an identity matrix.     $\blacksquare$

\vspace{8mm}
We now present a criterion for nonadditiveness of quantum codes. First a useful
notation. For a subset $\cal C$ of $\{ 0,1\}^n$ let
\[ {\cal T}({\cal C})=\left\{\, x\in\{0,1\}^n : x+{\cal C}\subseteq {\cal C}\,\right\} .\]
If $\cal C$ is a binary {\em linear} code then ${\cal T}({\cal C})={\cal C}$.

\begin{th}
Suppose that the quantum code $\cal Q$ of dimension $2^\ell$
is real and contains $\ket{\cal C}$, where
$\cal C$ is an $(n,K,d)$ binary code with $d>\lceil \log_2 K\rceil$. If the
identity operator is the only unitary operator in the stabilizer of $\cal Q$
and $2^{n-\ell}>|{\cal T}({\cal C})|$ then $\cal Q$ is nonadditive.
\label{nonaddtheo}
\end{th}

{\bf Proof.} Suppose, by contradiction, that $\cal Q$ is equivalent to additive 
code ${\cal Q}'$ via the transversal unitary operator $\cal U$ which mapps ${\cal Q}'$
on $\cal Q$. Let $\Gamma$ be the image of $\mbox{St}({\cal Q}')$ under $\cal U$.
Define $\Delta\subseteq \{0,1\}^n$ as in (ii) of Theorem \ref{equiadditheo}.  
Then $\Delta\subseteq {\cal T}({\cal C})$. Thus
\[ 2^{n-\ell}=|\mbox{St}({\cal Q}')| \leq |\Delta|\leq |{\cal T}({\cal C})| , \]
which contradicts the assumption of the theorem. $\blacksquare$

\vspace{8mm}
When the binary code $\cal C$ in the above theorem is linear we can formulate the theorem
as follows.

\begin{co}
Suppose that the quantum code $\cal Q$ of dimension $2^\ell$
is real and contains $\ket{\cal C}$, where
$\cal C$ is a linear $[n,k,d]$  code with $d>k$. If ${\mathrm St}({\cal Q})=\{I\}$
and $n>k+\ell$ then $\cal Q$ is nonadditive.
\label{linco}
\end{co}

Finally, we fomulate a criterion that guarantees strongly nonadditiveness
of quantum codes.

\begin{th}
Suppose that the qauntum code $\cal Q$ is real and it contains $\ket{\cal C}$ 
where $\cal C$ is an $(n,K,d)$ binary code with $d>\lceil\log_2 K\rceil$.
If $\mbox{\em St}({\cal Q})=\{ I\}$ and $GS({\cal Q})$ does not contain any 
operator of the form $X_\alpha T$, where $\alpha\neq{\bf 0}$ and $T$ 
is of the form {\em (\ref{tequ})}, then $\cal Q$ is strongly nonadditive.
\label{nonaddco}
\end{th}

{\bf Proof.} Suppose, by contradiction, that ${\cal Q}\subseteq {\cal Q}_1$
and ${\cal Q}_1\neq \cc^{2^n}$ is equivalent to an additive
code ${\cal Q}'$ with $\mbox{St}({\cal Q}')\neq \{I\}$.
Then, by Theorem \ref{equiadditheo}, any nontrivial stabilizer 
$\varphi$ of ${\cal Q}'$ defines an operator ${\cal V}=v_1\otimes\cdots\otimes v_n$ 
in $GS({\cal Q}_1)\subseteq GS({\cal Q})$, where $v_j=I$ or it is of the form 
(\ref{vmatrix}) or 
(\ref{umumatrix}). If all $v_j$ have real matrices, then ${\cal V}\neq I$ and 
${\cal V}\in\mbox{St}({\cal Q})$, which is impossible. If at least one of $v_j$
has a complex matrix, then $\cal V$ is of the form $X_\alpha T$ with $\alpha\neq
{\bf 0}$, which is again impossible.      $\blacksquare$

\subsection{Construction of nonadditive codes}
\label{nonaddsubsection}

\subsubsection{Examples of nonadditive codes}

Now we show that there is an infinite family of nonadditive quantum 
error--correcting codes.  These codes are constructed following the scheme 
similar to the one described in Theorem
2.4 of \cite{vra}.  Consider an $[n,k]$ binary code $\cal C$ such that
$\mbox{dist}({\cal C})$ and $\mbox{dist}({\cal C}^\perp)$ are both at least 
$d_0$ ($\cal C$ needs not to be a weakly self--dual code). 

First we define a function $\tau :{\cal C}\longrightarrow \{ 0,1\}^n$ such that
for $c,c'\in{\cal C}$ and $c\neq c'$ we have $\tau (c)+\tau (c')\not \in 
{\cal C}^\perp$. This means $\tau (c)$ and $\tau (c')$ are in different cosets
of ${\cal C}^\perp$ in $\{ 0,1\}^n$, for $c\neq c'$. Since there are $2^k$
different cosets, such mapping $\tau$ always can be defined. 
 
Fix $d\leq d_0$, and let $\cal E$ be the set
of binary vectors of length $n$ with weight $\leq d-1$.
Consider a subset $R=\{\, a_0,a_1,\ldots , a_m\, \}$ of $\{
0,1\}^n$ such that $a_0={\bf 0}$ and $a_j$ is not of the form $c+a_i+e$, for
$c\in {\cal C}$, $1\leq i\leq j-1$, and $e\in {\cal E}$.  Then the vectors
\begin{equation}
  \ket{x_i}=\sum_{c\in{\cal C}}(-1)^{\tau (c)\cdot a_i}\ket{c+a_i} 
\label{xidef}
\end{equation}
form a basis for a quantum code with distance $d$.  To prove this, we show that
$\langle x_i\mid X_\alpha Z_\beta \mid x_j\rangle=0$, for $0<\mbox{wt}
(\alpha\cup\beta)<d$. The case $\alpha\neq {\bf 0}$ or $i\neq j$ is
straightforward. So we only consider the case $\alpha={\bf 0}$ and $i=j$.
Then for $0<\mbox{wt}(\beta)<d$ we have
\begin{eqnarray*}
 \langle x_i\mid Z_\beta \mid x_i\rangle 
  & = & \left\langle \sum_{c\in{\cal C}}(-1)^{\tau(c)\cdot a_i}\ket{c+a_i}
           \left | \sum_{c\in{\cal C}}(-1)^{\tau(c)\cdot a_i+(c+a_i)\cdot
          \beta}\ket{c+a_i} \right . \right\rangle \\
  & = & (-1)^{a_i\cdot\beta}\sum_{c\in{\cal C}}(-1)^{c\cdot\beta} \\
  & = & 0.
\end{eqnarray*}
The last equality follows from the fact that $\mbox{dist}({\cal C}^\perp)
\geq d$, so $\beta\not\in{\cal C}^\perp$.

\begin{lm} 
In the above construction, suppose that 
\begin{equation}
              (n-1)2^k\sum_{i=0}^{d-1}{n\choose i} < 2^{n-1} .
\label{nkdcond}
\end{equation}
Then it is possible to choose $n$ linearly independent vectors $a_1,a_2,\ldots ,a_n$
so that the $((\, n,n+1,d\, ))$ quantum code $\cal Q$ with the basis 
$\ket{x_0},\ket{x_1},\ldots ,\ket{x_n}$ (each $\ket{x_i}$ is defined
by  {\em (\ref{xidef})})  has trivial stabilizer,
i.e.,  $\mbox{\em St}({\cal Q})=\{ I\}$.  
\label{nonadditivetheo} 
\end{lm}

{\bf Proof.}  
Suppose that the vectors $a_0,a_1,\ldots ,a_m$ with the desired properties are 
chosen. Then it is possible to choose a vector $a_{m+1}$ such that $a_1,\ldots,
a_m,a_{m+1}$ are independent and $a_{m+1}$ is not of the form $c+a_i+e$
(for $c\in{\cal C}$, $1\leq i\leq m$, and $e\in{\cal E}$) if
$\displaystyle 2^m+m\cdot 2^k\cdot \sum_{i=0}^{d-1}{n\choose i} <2^n$. This shows
that it is possible to choose $n$ vector $a_1,\ldots ,a_n$ with the desired 
properties.

Now we show that the identity operator is the only member of the stabilizer of 
$\cal Q$. Suppose that $X_\alpha Z_\beta$ is in the stabilizer of $\cal Q$. Since 
\[ X_\alpha Z_\beta \ket{x_0}=\sum_{c\in{\cal
                                  C}}(-1)^{c\cdot \beta} \ket{c+\alpha} \]
should be equal to $\displaystyle \ket{x_0}=\sum_{c\in {\cal C}}\ket{c}$
it follows that $\alpha \in {\cal C}$ and $\beta \in {\cal
C}^\perp$.  Similarly, for every $1\leq i\leq n$ since 
\begin{eqnarray*}
X_\alpha Z_\beta \ket{x_i} & = & \sum_{c\in{\cal C}}(-1)^{\tau (c)\cdot a_i+
                                   (c+a_i)\cdot \beta}\ket{c+a_i+\alpha} \\ 
                           & = & \sum_{c\in{\cal C}}(-1)^{\tau (c+\alpha)\cdot 
                                   a_i+(c+a_i+\alpha)\cdot \beta}\ket{c+a_i} \\ 
                           & = & \sum_{c\in {\cal C}} (-1)^{(\tau(c+\alpha)+
                                   \beta ) \cdot a_i}\ket{c+a_i}  
\end{eqnarray*} 
should be equal to
\[  \ket{x_i}=\sum_{c\in {\cal C}}(-1)^{\tau (c)\cdot a_i}\ket{c+a_i} , \] 
it follows that $a_i\cdot (\tau (c)+\tau (c+\alpha )+\beta )=0$, for every
$1\leq i\leq n$. Since $a_i$'s are independent, therefore 
$\tau (c)+\tau (c+\alpha)=\beta\in {\cal C}^\perp$, 
hence $\alpha ={\bf 0}$. Now the conditions $a_i\cdot \beta =0$ (for $1\leq 
i\leq n$) imply $\beta ={\bf 0}$. $\blacksquare$

\begin{th}
Suppose that $\cal C$ is an $[n,k,d_0]$ binary linear code such that $d_0>k$ and
$\mbox{dist}({\cal C})$ and $\mbox{dist}({\cal C}^\perp)$ are at least
$d$. Morover, suppose that $n$, $k$ and $d$ satisfy {\em (\ref{nkdcond})}. 
Let $\ell$ be the greatest integer such that $2^\ell\leq 
2^{n-k} / \sum_{i=0}^{d-1}{n\choose i}$. Suppose that $k+\ell < n$.
Then there is a an $((n,2^\ell,d))$ nonadditive code.
\end{th}

{\bf Proof.} Consider the $((n,n+1,d))$ code
${\cal Q}_0$ constructed in the previous lemma. Then by Theorem 4.2 of
\cite{vra} it is possible to add at least $2^\ell-(n+1)$ more vectors to
${\cal Q}_0$ to build an $((n,2^\ell,d))$ code $\cal Q$, which is, by 
Corollary \ref{linco}, nonadditive. $\blacksquare$

\vspace{8mm}

As an application we show that there are $((n,\lfloor 2^{n-1}/(n+1)\rfloor ,2))$
nonadditive codes, for every $n\geq 8$. Consider  the $[n,1,n]$
binary code ${\cal C}=\{ {\bf 0},{\bf 1}\}$. Then ${\cal C}^\perp$ is
consists of all even weight vectors in $\{ 0,1\}^n$, so it is an $[n,n-1,2]$
code. The condition (\ref{nkdcond}) satisfies if $n\geq 8$.
Then by applying the above theorem (for $k=1$ and 
$\ell=\lceil n-1-\log_2(n+1)\rceil$) we get the desired code.
Other classes of binary codes for which the minimum distance
of the code and its dual are known (such as Hamming codes and Reed--Muller 
codes) can be used to get nonadditive codes with different parameters. 

Finally, we show that the nonadditive codes are almost as good as
Calderbank--Shor--Steane (CSS) codes,
at least in the case that the dimension of code is large enough. The construction
of CSS codes was explained in the beginning of Section \ref{structure}. 

To utilize the CSS codes for constructing nonadditive codes, we must modify them such that
the new codes have trivial stabilizer. Let $\cal Q$ be an $[[n,n-2k,d]]$ CCS code based
on the weakly self--dual $[n,k]$ code $\cal C$ with  $\mbox{dist}({\cal C}^\perp)\geq d$.
Consider the basis for $\cal Q$ consists of vectors 
$\displaystyle\ket{x_a}=\sum_{c\in{\cal C}}\ket{c+a}$, for $a\in{\cal C}^\perp / {\cal C}$.
Also consider the function $\tau\colon {\cal C}\longrightarrow \{0,1\}^n$ defined at the
beginning of this section. We define the quantum code $\wh{\cal Q}$ with basis
\begin{equation}
  \ket{y_a}=\sum_{c\in{\cal C}}(-1)^{\tau(c)\cdot a}\ket{c+a} , 
\label{yadef}
\end{equation}
for $a\in{\cal C}^\perp / {\cal C}$. Then it is easy to check that $\wh{\cal Q}$
is also an $[[n,n-2k,d]]$ code. 

\begin{th}
Suppose that $\cal C$ is an $[n,k,d_0]$ weakly self--dual binary code, and ${\cal C}^\perp$
is an $[n,n-k,d_1]$ code. Assume $d_0\geq k$ and $2^{n-2k-1}>n-k-1$ (for example 
it is enough that $k<(n-\log_2n)/2$). For any $d\leq d_1$ that staisfies
\begin{equation}
 \left ( 2^{n-k}+(k-1)2^{k}\right ) \sum_{i=0}^{d-1}{n\choose i} < 2^{n-1} , 
\label{dcond}
\end{equation}
we have an $((n,2^{n-2k},d))$ nonadditive code.
\label{cssnonaddtheo}
\end{th}

{\bf Proof.} Let ${\cal Q}_0$ be the $[[n,n-2k,d]]$ CSS code based on $\cal C$, and let
$\wh{{\cal Q}_0}$ be the quantum code obtained from ${\cal Q}_0$ as described in the above.
We can choose independent vectors $a_1,\ldots ,a_{n-k}$ in 
${\cal C}^\perp$ such that $a_i$'s belong to different cosets of $\cal C$
in ${\cal C}^\perp$. This is possible because $2^{n-2k-1} > n-k-1$.
We consider $\ket{y_{a_1}},\ldots ,\ket{y_{a_{n-k}}}$ (defined by (\ref{yadef}))
as vectors in $\wh{{\cal Q}_0}$. Then we
choose vectors $a_{n-k+1},\ldots ,a_n$ such that $a_1,\ldots ,a_n$ are $n$
independent vectors, and ${\cal Q}'=\wh{{\cal Q}_0}\cup 
\left\{ \ket{x_{a_{n-k+1}}},\ldots ,\ket{x_{a_n}}\right\}$, 
is an $(( n,2^{n-2k}+k,d))$ code. The inequality 
(\ref{dcond}) implies that it is possible to choose $a_{n-k+1},\ldots ,a_n$
with the desired properties. Then the proof of Lemma 
\ref{nonadditivetheo} shows that $\mbox{St}({\cal Q}')=\{ I\}$

Let $\cal Q$ be the quantum code obtained from ${\cal Q}'$ by removing any
$k$ vectors except $\ket{y_{a_i}}$, $i=1,\ldots ,n$. 
Then $\mbox{St}({\cal Q})=\{ I\}$ (because $\cal Q$ contains the $\ket{y_{a_i}}$, 
$i=1,\ldots ,n$).
So, by Corollary \ref{linco} with 
$\ell=n-2k$, $\cal Q$ is nonadditive. $\blacksquare$ 
                                                       
\vspace{8mm}
To show that there are weakly self--dual codes $\cal C$ that satisfy the 
requirements of the above theorem, apply the greedy method used in 
classical coding theory (see \cite{macwilliams}, Chap. 17). The same 
method is used in \cite{calderbank} to prove the existence of CSS codes
meeting the Gilbert--Varshamov bound.

Suppose that $n$ is even.
Let $\Phi_{n,k}$ be the set of all $[n,k]$ weakly self--dual codes; and
$\Phi_{n,k}'$ be the set of all codes ${\cal C}^\perp$ where $\cal C$ is 
in $\Phi_{n,k}$. Let $\varphi=|\Phi_{n,k}|=|\Phi_{n,k}'|$. In 
\cite{macwilliams72} (see also \cite{macwilliams} p. 630) it is shown that
every {\em nonzero} vector $v$ with even weight
belongs to exactly $\sigma_{n,k}$ codes in 
$\Phi_{n,k}$, where the number $\sigma_{n,k}$ does not depend on the vector
$v$. It is also shown in \cite{calderbank} that every even--weight vector 
$v\not \in \{ {\bf 0},{\bf 1}\}$  belongs
to exaclty $\sigma_{n,k}'$ codes in $\Phi_{n,k}'$. Then
\begin{eqnarray*}
   \left ( 2^{n-1}-1\right ) \sigma_{n,k}  & = & \left ( 2^k-1\right ) \varphi , \\
   \left ( 2^{n-1}-2\right ) \sigma_{n,k}' & = & \left ( 2^{n-k}-2\right ) \varphi .
\end{eqnarray*}
Then the number of codes in $\Phi_{n,k}'$ with minimum distance $\leq d$ is
at most
\begin{eqnarray*}
   \sum_{j=0}^d{n\choose j}\sigma_{n,k}' & \leq & 2^{H_2(d/n)n}\sigma_{n,k}' \\
					 & \leq & 2^{H_2(d/n)n-k+1}\varphi ,
\end{eqnarray*}
where $H_2$ is the binary entropy function $H_2(t)=-t\log_2t-(1-t)\log_2(1-t)$.
Let $k=\lceil H_2(d/n)n\rceil +3$, then more than ${3\over 4}$ of the codes in 
$\Phi_{n,k}'$ have minimum distance greater than $d$. 
Now in the class $\Phi_{n,k}$, for the value of $d_1$ such that $k\leq d_1$ and
$k\leq n-H_2(d_1/n)n-2$, it follows that at most half of the codes in $\Phi_{n,k}$
have minimum distance $\leq d_1$; because the number of codes in $\Phi_{n,k}$
that contain a codeword of weight $<d_1$ is at most
\begin{eqnarray*}
\sum_{j=0}^{d_1}{n\choose j}\sigma_{n,k} & \leq & 2^{H_2(d_1/n)n}\sigma_{n,k} \\
				         & \leq & 2^{H_2(d_1/n)n+k-n+1}\varphi \\
                                         & \leq & 2^{-1} \varphi .
\end{eqnarray*}

Let $d=\alpha n$ and $d_1=\beta n$. The above conditions on $k$, $d$ and $d_1$ satisfy
if $H_2(\alpha)<\beta$ and $H_2(\alpha) < 1-H_2(\beta)$. We show that there are $\alpha$
and $\beta$ that satisfy these inequalities. Choose $\delta _1,\delta_2<{1\over 2}$ such that
$H_2(\delta_1)={1\over 2}$ and $H_2(\delta_2)=\delta_1$. Choose $\alpha<\delta_2$. Then
$H_2(\alpha)<\delta_1$. Choose $\beta$ such that $H_2(\alpha)<\beta<\delta_1$.
Then $1-H_2(\beta)>1-H_2(\delta_1)={1\over 2}>H_2(\alpha)$.
So let
$\alpha < H_2^{-1}(H_2^{-1}(1/2))\approx 0.0146$, where $H_2^{-1}$ is the inverse of
the entropy function. With this bound on $d$, we showed that that there is a weakly
self--dual $[n,k,d_1]$ code $\cal C$ such that $d_1 > k$ and ${\cal C}^\perp$ is an 
$[n,k,d]$ code with $k/n \approx H_2(d/n)$. Note that the condition (\ref{dcond}) also
holds, because the left--hand side of this inequality is at most 
$2^{n-k+H_2(d/n)n+1}$, which for the chosen value for $k$, is less than $2^{n-2}$.
So we have shown the following asymptotic bound.

\begin{th}
For $d<\lambda n$, where $\lambda =H_2^{-1}(H_2^{-1}(1/2))$, there are nonadditive
$((n,2^k,d))$ quantum codes with rate $k/n \geq 1-2H_2(d/n)$.
\end{th}

\subsubsection{A strongly nonadditive code}

In this section we provide an example of a strongly nonadditive quantum 
error--correcting code. This is an $((11,2,3))$  strongly nonadditive code.

Consider the (Paley type) Hadamard matrix of order 12 (see, e.g.,
\cite{macwilliams}, p.  48).  Delete the all--1 column and replace $-1$ by 1 and
$+1$ by 0.  The result is the following matrix 
\[ H=\left [ \matrix{ 
      0 & 0 & 0 & 0 & 0 & 0 & 0 & 0 & 0 & 0 & 0 \cr 
      1 & 0 & 1 & 0 & 0 & 0 & 1 & 1 & 1 & 0 & 1 \cr
      1 & 1 & 0 & 1 & 0 & 0 & 0 & 1 & 1 & 1 & 0 \cr
      0 & 1 & 1 & 0 & 1 & 0 & 0 & 0 & 1 & 1 & 1 \cr 
      1 & 0 & 1 & 1 & 0 & 1 & 0 & 0 & 0 & 1 & 1 \cr 
      1 & 1 & 0 & 1 & 1 & 0 & 1 & 0 & 0 & 0 & 1 \cr 
      1 & 1 & 1 & 0 & 1 & 1 & 0 & 1 & 0 & 0 & 0 \cr 
      0 & 1 & 1 & 1 & 0 & 1 & 1 & 0 & 1 & 0 & 0 \cr 
      0 & 0 & 1 & 1 & 1 & 0 & 1 & 1 & 0 & 1 & 0 \cr 
      0 & 0 & 0 & 1 & 1 & 1 & 0 & 1 & 1 & 0 & 1 \cr 
      1 & 0 & 0 & 0 & 1 & 1 & 1 & 0 & 1 & 1 & 0 \cr 
      0 & 1 & 0 & 0 & 0 & 1 & 1 & 1 & 0 & 1 & 1 \cr } 
\right ] .  \]
We denote the $i^{\mbox{\scriptsize th}}$ row of $H$ by $r_i$. The set
${\cal C}=\{\, r_i : 1\leq i\leq 12\,\}$ is an $(11,12,6)$ code. Then a basis
for the desired quantum code consists of the following two vectors:
\pagebreak
\begin{eqnarray*} 
           \ket{0_L} & = & \sum_{i=1}^{12} \ket{r_i} , \\ 
           \ket{1_L} & = & \sum_{i=1}^{12} \ket{\mbox{\bf 1}+r_i} , \\ 
\end{eqnarray*} 
where {\bf 1} is the all--1 vector of length 11.    
We claim these vectors are basis for an $((11,2,3))$ quantum code.  We have to show that
\begin{eqnarray} 
\left \langle 0_L\, \right | \,X_\alpha Z_\beta \, \left | \, 0_L \right \rangle 
                                & = & 0 , \label{eq1}\\ 
\left\langle 1_L\, \right | \, X_\alpha Z_\beta \, \left | \, 1_L \right \rangle 
                                & = & 0 , \label{eq2} \\ 
\left \langle 0_L\, \right | \, X_\alpha Z_\beta \, \left |\, 1_L \right \rangle 
                                & = & 0 ,  \label{eq3} 
\end{eqnarray} 
for every
$\alpha,\beta\in\{0,1\}^{11}$ such that $1\leq \mbox{wt}(\alpha\cup \beta )\leq
2$.  First note that that the distance of any two distinct vectors in the set 
\[ \left \{\, r_i :  1\leq i\leq 12\, \right \} \cup \left \{\, \mbox{\bf 1}+
r_i : 1\leq i\leq 12\, \right \} \] 
is at least 5.  Thus if $1\leq
\mbox{wt}(\alpha)\leq 4$ then all conditions (\ref{eq1})--(\ref{eq3}) hold.  Now
suppose that $\alpha = \mbox{\bf 0}$.  Then (\ref{eq3}) trivially holds.
To see that (\ref{eq1}) and (\ref{eq2}) hold it is enough to note that
if $1\leq
\mbox{wt}(\beta )\leq 2$ then $r_i\cdot \beta =1$ for exactly 6 values of $i$.
This completes the proof that $\left \{\, \ket{0_L}, \ket{1_L}\,
\right \}$ is a basis for an $((11,2,3))$ quantum error--correcting
code.

To show that this code is nonadditive, let $\vph=(-1)^\lambda X_\alpha
Z_\beta$ be any operator in the stabilizer of this code.  Since $\vph
\ket{0_L}=\ket{0_L}$ and $\vph \ket{r_1}=\ket{\alpha}$, hence $\lambda =0$ and 
$\alpha$ should be one of $r_i$'s.  Then we should have $\alpha
=r_1=\mbox{\bf 0}$, because for every $r_i$, $i\neq 1$, there is some
$j$ such that $r_i+r_j$ is not equal to any $r_k$.  Therefore, $\vph =
Z_\beta$.  Then
\[ Z_\beta\ket{0_L}=\sum_{i=1}^{12}(-1)^{r_i\cdot 
\beta}\ket{r_i}=\sum_{i=1}^{12}\ket{r_i} \] 
implies that $r_i\cdot \beta =0$, for every $i$.  But the set $\left
\{\, r_i : 1\leq i\leq 12\, \right \}$ has rank 11, so $\beta
=\mbox{\bf 0}$.  This shows that the identity operator is the only
operator in the stabilizer of this code. Finally, suppose that $X_\alpha T$
is in the generalized stabilizer of this code, where the operator $T$
is of the form (\ref{tequ}). Note that the operator $T$ only effects
the phases of the states, so the above argument also implies $\alpha
={\bf 0}$. Now Theorem \ref{nonaddco} implies that this code is
strongly nonadditive.

\section{Concluding Remarks}

We gave a characterization of additive quantum codes, and showed that 
there are nonadditive codes with different minimum distances.  We showed
that nonadditive codes that correct $t$ errors can reach the asymptotic
rate $R\geq 1-2H_2(2t/n)$.  We introduced the notion of strongly
nonadditive codes, and gave an example of such codes.  It would be
interesting to find more examples of such codes.  We conjecture that
the nonadditive codes constructed in Secition~4.2.1 are also strongly
nonadditive codes.


\begin{thebibliography}{11}

\bibitem{bdsw} 
C.  H.  Bennett, D.  P.  DiVincenzo, J.  A.  Smolin and W.  K. Wootters, 
``Mixed state entanglement and quantum error correction,'' {\em Phys.
Rev.  A}, Vol.  54, No.  5, pp.  3824--3851 (1996).

\bibitem{grasslbeth} 
M.  Grassl and Th.  Beth, ``A note on non--additive quantum
codes,'' LANL e--print quant--ph/97030126.

\bibitem{orthogeo} 
A.  R.  Calderbank, E.  M.  Rains, P.  W.  Shor and N.  J. A.  Sloane, 
``Quantum error correction and orthogonal geometry,'' LANL e--print
quant--ph/9605005.

\bibitem{gf4} 
A.  R.  Calderbank, E.  M.  Rains, P.  W.  Shor and N.  J.  A. Sloane, 
``Quantum error correction via codes over GF(4),'' LANL e--print
quant--ph/9608006.

\bibitem{calderbank} 
A.  R.  Calderbank and P.  W.  Shor, ``Good quantum error--correcting 
codes exit,'' {\em Phys.  Rev.  A}, Vol.  54, No.  2, pp. 1098--1105 (1996).

\bibitem{cleve} 
R.  Cleve, ``Quantum stabilizer codes and classical linear
codes,'' LANL e--print quant--ph/9612048.

\bibitem{gottesman} 
D.  Gottesman, ``A class of quantum error--correcting codes
saturating the quantum Hamming bound,'' {\em Phys.  Rev. A}, Vol.  54, No.  3,
pp.  1862--8168 (1996).

\bibitem{knilllaf} 
E.  Knill and R.  Laflamme, ``A theory of quantum
error--correcting codes,'' LANL e--print quant--ph/9604034.

\bibitem{macwilliams72} F. J. MacWilliams, N. J. Sloane and J. P. Thompson,
``Good self dual codes exist,'' {\em Discrete Math.}, vol. 3, pp. 153--162
(1972).

\bibitem{macwilliams} 
F.  J.  MacWilliams and N.  J.  A.  Sloane, {\em The
Theory of Error Correcting Codes}, North--Holland, New York, 1977.

\bibitem{rains_shadow}
E. M. Rains, ``Quantum shadow enumerators,''
LANL e--print quant--ph/9611001.

\bibitem{rains_d2}
E. M. Rains, ``Quantum codes of minimum distance two,''
LANL e--print quant--ph/9704043.

\bibitem{nonadditive} 
E.  M.  Rains, R.  H.  Hardin, P.  Shor and N.  J.  A. Sloane, 
``A nonadditive quantum code,'' LANL e--print quant--ph/9703002.

\bibitem{steane} 
A.  M.  Steane, ``Error correcting codes in quantum theory,''
{\em Phys.  Rev.  Lett.}, Vol.  77, No.  5, pp.  793--797 (1996).

\bibitem{vra} 
F.  Vatan, V.  P.  Roychowdhury and M.  P.  Anantram, ``Spatially
correlated qubit errors and burst--correcting quantum codes,'' LANL e--print
quant--ph/9704019.

\end{thebibliography}
\end{document}